\documentclass[conference]{IEEEtran}
\usepackage{amsfonts}
\usepackage{cite}
\usepackage{url}
\usepackage{graphicx}

\begin{document}
\title{Internet Voting Protocol Based on Implicit Data Security}

\author{\IEEEauthorblockN{Abhishek Parakh}
\IEEEauthorblockA{Computer Science Department\\
Oklahoma State University\\
Stillwater, OK 74075\\
Email: parakh@cs.okstate.edu}
\and
\IEEEauthorblockN{Subhash Kak}
\IEEEauthorblockA{Computer Science Department\\
Oklahoma State University\\
Stillwater, OK 74075\\
Email: subhashk@cs.okstate.edu}
}

\maketitle

\begin{abstract}
This paper presents a new protocol for Internet voting based on implicit data security. This protocol allows recasting of votes, which permits a change of mind by voters either during the time window over which polling is open or during a shorter period over which recasting is permitted. The security of votes depends on multiple servers such that each vote is divided into partitions and these partitions are distributed among the servers, all of which need to be brought together to reconstruct the votes. Such a protocol has potential applications in bargaining and electronic commerce.
\end{abstract}

\section{Introduction}
Security in Internet voting is a basic research problem due to the opposite requirements of confidentiality and verifiability. Several cryptographic solutions have been proposed to meet these requirements, but their effectiveness is predicated on several assumptions. In general, security of voting and its trustworthiness depends on the honesty of numerous agents, human and virtual, and it is therefore easy to argue that universal verifiability and unconditional privacy cannot be simultaneously achieved. Nevertheless, we provide an effective solution based on implicit data security, which has several attractive features from the point of view of implementation.

An elaborate framework for data security based on implicit architecture is described in \cite{ref21}. Online data security has direct applications in Internet voting because each vote can be regarded as a voter's data that needs to be protected. One advantage of using vote (data) partitioning scheme in Internet voting is that one does not require an encryption/decryption key and that the security is implicit in the partitions. None of the partitions reveal the vote when the election is open and only when all the partitions are brought together, after the closing of elections, can the votes be recreated.

Although several problematic issues remain with current approaches to electronic voting \cite{ref19}, Internet voting is generally expected to become widespread in the next few years and is already in use in Switzerland, Estonia, England, and Netherlands, and on experimental basis in the United States \cite{ref17}. It offers ease of access to senior citizens, disabled people, people who are traveling on the election-day, citizens living abroad but who are eligible to vote, soldiers serving abroad, and eliminates the hassle of obtaining an absentee ballot in advance. It also encourages larger participation by the younger generation that has become accustomed to online banking, online shopping, secure email transactions and secure online storage.

However, as in other network technologies, Internet voting is vulnerable to viruses, Trojan horses and denial of service attacks. A number of measures can be taken to nullify the consequence of such attacks \cite{ref18}, one of which is holding the elections over a period of many days, as in the case of Switzerland for two weeks \cite{ref14}, and in Estonia \cite{ref15} for a few days. More trust may be obtained \cite{ref10} if all the voters actually voted, which in practice is almost impossible to ensure.

When an election is open for a long period, an undecided voter may wish to change the vote as the election debate progresses during this period. If not allowed to do so (i.e. change the vote), the voter may tend to wait until the last minute to cast his vote and to arrive at a final decision. Avoiding ``last minute voting" over the Internet is desirable so as to decrease the chances of network attacks and disruptions aimed at the final ``rush hour."

Also, voters can accidentally vote for the wrong candidate (similar to errors in choosing wrong options in online bookings and purchases), instances of which were noticed in the California recall elections \cite{ref2, ref13}. Therefore, a voter may wish to correct his mistake by casting another ballot overwriting the previous one. As a result, it is essential that electronic voting include a provision to recast votes. None of the present schemes allows voters to do so.

\section{Previous Work}
A number of electronic voting schemes are described in the literature \cite{ref1, ref3, ref4, ref5, ref6, ref11, ref12, ref19}. An overview of the problems of Internet voting and implementation strategies may be found in \cite{ref8}. Convertible blind signatures \cite{ref16} have been applied on the ballots in \cite{ref1} to achieve ballot secrecy, and the intermediate results (fairness property) protected using secretly generated random numbers. However, it is not clear how the voter provides the authorities with this secret information after the election is closed. In practice the voter cannot be expected to go back to the election website in order to provide the decryption key. Complicated Mix-Nets are used to provide anonymization in \cite{ref3, ref4, ref5, ref11}, which we believe can be avoided using anonymous IDs that are untraceable to the voter's real identity.

Juels et al \cite{ref6} propose a practical framework for developing Internet voting schemes, but they assume an untappable channel during the registration phase which is not truly available in practice. Postal mail is suggested as an untappable channel, which is not true considering the number of identity thefts that occur due to its insecurity \cite{ref7}. Moreover, this scheme is not suitable for large scale elections due to its large overhead.

The voting scheme presented in \cite{ref12} and \cite{ref8} replaces e-cash with vote. It does not allow the voter to recast a ballot and aims at revealing voter's true identities in case of attempt at recast. To the authors' knowledge there does not exist any scheme that allows a ballot recast and presently all the intermediate election results are protected using encryption. The article aims at presenting a scheme that overcomes this limitation and does not require any explicit encryption of votes. In general an electronic voting scheme needs to satisfy the following properties \cite{ref1, ref3, ref5, ref6}:
\begin{enumerate}
\item Receipt freeness: The voter must not possess any proof of the cast ballot that can be used to convince a third party of the way in which the voter has voted.
\item Privacy: No one should be able to determine how the voter has voted.
\item Un-reusability: A voter can cast only a single vote.
\item Un-forge-ability: An ineligible voter cannot cast a ballot.
\item Fairness: Intermediate elections results must not be available as they provide incentive for fraud.
\item Distributed security: The security of the election process must rely upon and must be overseen by more than one authority. This has been traditionally achieved by numerous poll workers and helper agencies (NGOs), under the assumption that not all of them have been compromised at any given stage.
\end{enumerate}
The security of the system largely depends on the effectiveness of the distributed systems and multiple audit trails.

\section{Our Contribution}
Our contribution in this article is twofold. Firstly, we present a scheme that allows voters to recast ballots in case they wish to change their decision or have accidently voted for a wrong candidate. The proposed scheme achieves this by generating an anonymous ID at the time of registration and retrieving a set of authentic-signed ballots that can be reused. Each vote is tracked by the anonymous ID and only the latest cast ballot is taken into consideration. The scheme is implemented so that when a recast in intended, the voter need not contact the registration authority. A variation in which the voter may wish to contact (under certain circumstances) the registration authority while changing vote may also be implemented.

Secondly, we protect the intermediate results of election by using implicit data security. We assume that a voter has $k$ servers at his disposal and he votes by creating $k$ partitions of his ballot and sending them to the servers along with his anonymous ID. The implementation of this partitioning could be done in one of the two ways: (1) by an application residing on the voter's computer which sets up a secure connection with the servers that will record the partitions, and (2) At a primary server that, in turn, sends the partitions to other servers. In most cases, the second approach may be more trustworthy. All of these partitions need to be brought together to recreate the final vote during the counting phase. Security of the system is such that as long as one of the servers is honest any attempt to change a vote by manipulating $k-1$ partitions succeeds with a probability of $1/p$, where $p$ is a large prime. For example, if $p\approx2^{100}$, then the probability of changing a vote as desired, without detection, is $1/2^{100}$. Further, the probability of changing a vote into another valid vote, not necessarily the desired vote, can be achieved with a probability of $m/p$, where $m$ is the number of candidates contesting in the election.
\section{The Proposed Protocol}
The following notations are used in the description of the protocol.\\
$V_{id}$: the eligible voter's true identification. This may include a national ID number (Social Security Number in the US) and certain personal information similar to that provided by voters to a credit score reporting company over the Internet.\\
$r_{id}$: a random identification number that the voter generates and uses at the time of casting his ballot. This, when signed by the registration authority acts as voter's anonymous ID.\\
$C_j$: the $j^{th}$ candidate belonging to a set $C$, and the cardinality $|C|=m$.\\
$g^x$: the public key that the registration authority (RA) announces, where $x$ is the corresponding private key. Computations are performed modulo a prime $p$ where $g$ is a primitive root and public knowledge. Therefore, $m^x$ is the signature on a message $m$, where $1<m<(p-1)$.\\
${ballot}_j$: the ballot for the $j^{th}$ candidate. It is a randomly chosen number from the field $\mathbb{Z}_p$ and is publicly known. Hence, if a voter wishes to vote for a candidate he needs to submit this random number as the cast ballot.\\
${server}_j$: is the $j^{th}$ server from the set of servers $S$ and $|S|=k$.
\subsection{Registration Phase}
\begin{enumerate}
\item Voter generates a random identification number $r_{id}$, chosen uniformly from a set of 100 digit numbers.
\item Voter generates a blinding factor $b$ independently and uniformly and sends to the registration authority (RA): $r_{id}\cdot g^b\textrm{ mod }p$, $V_{id}$.
\item The registration authority determines if the voter is eligible to vote and determines to what precinct the voter belongs. Then the registration authority signs the voters blinded $r_{id}$ and sends him a set of signed ballots that he/she can use for voting. The registration authority replies with the following message:
    $({r_{id}\cdot g^b})^x\textrm{ mod }p$, $(ballot_{1})^{x}, (ballot_{2})^{x}, ..., (ballot_{m})^{x}\textrm{ mod }p$
\item The voter now un-blinds his anonymous $r_{id}$ as follows:
\begin{enumerate}
    \item Using RA's public key compute $(g^x)^b\textrm{ mod }p$ and then compute its multiplicative inverse.
    \item Multiply $({r_{id}\cdot g^b})^x\textrm{ mod }p$ with the multiplicative inverse computed in step a. and retrieve $({r_{id}})^x\textrm{ mod }p$.
\end{enumerate}
\item Voter now issues a zero-knowledge challenge/response protocol to verify RA's signatures on $r_{id}$ and the set of ballots he has received. If the signature is found invalid then the voter can launch a disavowal protocol. The proper formation of the ballots can be checked by the voter if he wishes.
\end{enumerate}
The voter at the end of registration phase has a valid signed $r_{id}$ and a set of signed ballots that can be used to vote. The voter may proceed to cast his ballot as follows.
\subsection{Voting Phase}
\begin{enumerate}
\item Voter contacts his assigned online polling booth using a secure shell (SSL) connection over the Internet and sends: $(r_{id})^x \textrm{ mod }p$, $r_{id}$.
\item The polling booth verifies the validity of $r_{id}$ using either of the following approaches:
\begin{enumerate}
\item Each online polling booth may have a copy of $x$, in which case checking for the signature is trivial.
\item Alternatively, the online polling booth may conduct a zero-knowledge challenge/response protocol with the RA to validate the signature.
\end{enumerate}
    In case a collision is detected, the voter may need to contact the RA again and obtain another singed $r_{id}$.
\item If the signature is found valid the online polling booth stores the voter's $r_{id}$ and provides the voter with a secure session key that he can use to cast his ballot by contacting the $k$ voting servers.
\item The voter chooses the ballot corresponding to the candidate that he wishes to vote for and divides the ballot into $k$ partitions and sends one partition to each of the $k$ servers. In order to partition the ballot, the voter uses the following procedure:
\begin{enumerate}
\item The voter generates an equation of degree $k$ such that its coefficients belong to $\mathbb{Z}_p$. For example, $x^k+a_{k-1}x^{k-1}+...+a_1+ballot\equiv0 \textrm{ mod }p$, where $a_i$, $ballot\in\mathbb{Z}_p$.
\item He then computes the $k$ roots of the equation such that, $(x-r_1)(x-r_2)...(x-r_k)\equiv0\textrm{ mod }p$. It is to noted that roots may not exist for any arbitrary equation and the coefficients may need to be adjusted. Also, $r_1\cdot r_2\cdot...\cdot r_k\equiv ballot \textrm{ mod }p$.
\item The voter now stores these roots on different servers as partitions.
\end{enumerate}
\end{enumerate}
\subsection{Counting Phase}
\begin{enumerate}
\item All the servers pool together the partitions of the vote and recreate the secret by multiplying them in the finite field $\mathbb{Z}_p$.
\item The signature is verified on the ballots and if found valid the ballot is included into the final tally.
\end{enumerate}
\subsection{Re-voting}
If a voter wishes to change his vote, he may do so by contacting the online polling booth using his $r_{id}$ to authenticate himself and simply choose another ballot from the set of signed ballots and divide it into partitions and distribute them to other servers. The servers will overwrite his previously cast ballot partitions if any.
\begin{figure}
\centering
\includegraphics[width=3.3in, height=2.1in]{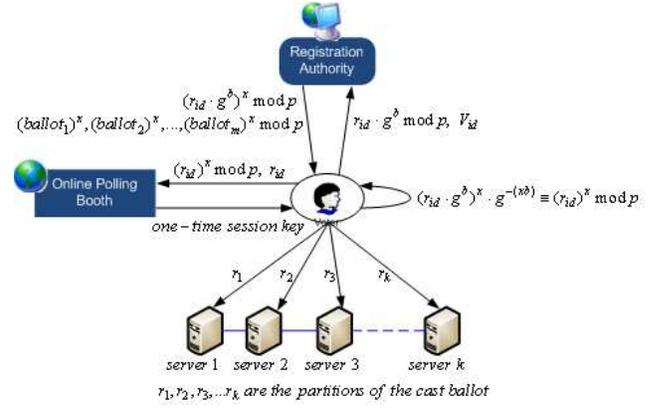}
\caption{Illustration of proposed voting protocol}
\label{fig_sim}
\end{figure}
\section{Analysis of the protocol}
\newtheorem{theorem}{Theorem}
\begin{theorem}
We assert the following regarding the proposed protocol:
\begin{enumerate}
\item Even if $k-1$ servers collude, they do not gain any information about the cast ballot.
\item If $k-1$ colluding servers wish to change the cast ballot into another desired ballot, then this can be achieved only with a probability of $1/p$.
\item The probability of $k-1$ colluding servers changing the cast ballot to another valid ballot is $m/p$.
\end{enumerate}
\end{theorem}
\begin{IEEEproof}
The proof of the assertions is presented below.
\begin{enumerate}
\item One of the ways to create the partitions would be that the voter chooses $k-1$ roots at random and computes the $k^{th}$ root using $r_k=r_1\cdot...\cdot r_{k-1}\cdot ballot \textrm{ mod } p$.
    Since, the colluding servers do not know that cast ballot, knowledge of all but one root leaves the last root undetermined and vice versa. Without knowing the last root, the cast ballot remains undetermined.
\item All the operations are performed in the field $\mathbb{Z}_p$ such that each candidate's ballot is a randomly chosen number (known to public) from the field. Further a signed ballot is again an integer within the field. Therefore there exist $m$ integers in the field that are valid singed ballots. From the part 1 of the proof, we notice that even the knowledge of $k-1$ roots does not reveal the ballot. Consequently, if the colluding servers wish to change the cast ballot into their desired ballot by only manipulating $k-1$ (or less) partitions then the best attempt they can make is by fixing $k-2$ partitions and changing the value of $(k-1)^{st}$ partition so that the final result is the desired ballot. However, since they do not know the value of the $k^{th}$ partition (which is also fixed) and neither do they know the value of the already cast ballot, the colluding servers can achieve their goal only with a probability of $1/p$.
\item It follows from part 2 of the proof, that since there are $m$ valid ballots, manipulating $k-1$ (or less) partitions will result in a valid singed ballot with a probability of $m/p$.
\end{enumerate}
\end{IEEEproof}
\begin{theorem}
The knowledge of $k-1$ partitions is equivalent to the knowledge of $k-i$ partitions, $1\leq i\leq k-2$, i.e. a server does not gain any information about the ballot by colluding with other servers as long as one at least of the servers does not collude and manipulation of $k-1$ ballots is equivalent to manipulation of $k-i$ partitions. In essence, no advantage is gained by collusion (if at least one of the servers remains honest).
\end{theorem}
\begin{IEEEproof}
Assume that the partitions are created as described in part 1 of Theorem 1 and that the servers do not know the value of the original cast ballot (which is practically the case). Therefore if $k-1$ partitions are fixed to a constant, the entire field $\mathbb{Z}_p$ can be `swept' by changing the value of only the $k^{th}$ partition. Therefore, knowledge of one or more partitions as long as the $k^{th}$ partition is kept secret does not help in manipulating the reconstruction of the ballot and servers do not gain any advantage by collusion. Further, any of the colluding servers can sweep the entire field by changing only its partition and does not need the other servers to change any of their partitions. Hence, collusion is of no use and the probabilities of creating a valid/desired output remains unchanged by collusion.
\end{IEEEproof}
Other properties of the protocol are analyzed below:
\begin{enumerate}
\item Receipt freeness: The proposed protocol is receipt free and it is assumed that at least one of the servers will be honest (an assumption that is present in the mix-net schemes as well) providing assurance to the voter that his vote will be counted as cast and any discrepancy will be detected with a very high probability.
\item Un-forge-ability: An ineligible voter cannot cast a vote because two parts constitute a vote, the signed ballot and the certified $r_{id}$. For an ineligible voter to cast a ballot, he must generate a random ID and forge the signature of the registration authority, i.e. solve for $x$, given $g^x\textrm{ mod }p$, $g$ and $p$. This is equivalent to solving the discrete log problem and computationally infeasible.
    Leakage of information about the signature exponent is avoided by using a zero-knowledge challenge and response protocol for signature verification.
    Note that even if an eligible voter provides an ineligible voter with a copy of signed ballots, the ineligible voter will still need a valid signed anonymous ID to cast a vote, which cannot be generated without registering.
\item Un-reusability: Since every eligible voter is issued only one $r_{id}$, he cannot use it to cast multiple votes using it because if he does so the servers will simply overwrite the previous cast ballot for that ID. Furthermore, forging a new $r_{id}$ is ruled out by property 2 above.
\item Fairness: The ballot cast is divided into partitions during the voting phase and unless all the partitions are known, all the ballots remain equally likely. This is established by the above theorems.
\end{enumerate}
\section{Conclusion}
We have presented a new protocol for Internet voting in which the security of the cast ballots depends on numerous servers and the fairness property is satisfied by the use of a ballot partitioning scheme. No encryption/decryption key is used and there is no explicit encryption of the votes. The partitions of the ballot provide implicit security. Variations of this protocol may be used in a variety of applications in bargaining and electronic commerce.

\end{document}